%% file: Manuscript.tex
\documentclass[reprint,12pt]{JASA}
\newcommand\myscale{1}

\begin{document}

\title[JASA/OMOQSE for TSM]{Deep Learning-Based Single-Ended Objective Quality Measures for Time-Scale Modified Audio}


\author{Timothy Roberts}
\author{Aaron Nicolson}
\author{Kuldip K. Paliwal}
\affiliation{Signal Processing Laboratory, Griffith University, 170 Kessels Road, Nathan, QLD 4111, Australia}
\email{timothy.roberts@griffithuni.edu.au}

\preprint{Roberts, Griffith University}

\date{\today} 

\begin{abstract}
Objective evaluation of audio processed with Time-Scale Modification (TSM) is seeing a resurgence of interest.  Recently, a labelled time-scaled audio dataset was used to train an objective measure for TSM evaluation.  This DE measure was an extension of Perceptual Evaluation of Audio Quality, and required reference and test signals.  In this paper, two single-ended objective quality measures for time-scaled audio are proposed that do not require a reference signal.  Data driven features are created by either a convolutional neural network (CNN) or a bidirectional gated recurrent unit (BGRU) network and fed to a fully-connected network to predict subjective mean opinion scores.  The proposed CNN and BGRU measures achieve an average Root Mean Squared Error of 0.608 and 0.576, and a mean Pearson correlation of 0.771 and 0.794, respectively.  The proposed measures are used to evaluate TSM algorithms, and comparisons are provided for 16 TSM implementations.  The objective measure is available at https://www.github.com/zygurt/TSM.


\end{abstract}


\maketitle



\section{Introduction}
\label{sec:1}
Time-Scale Modification (TSM) aims to manipulate the temporal domain of a signal independent of  pitch and timbre.  The time-scale ratio ($\beta$) denotes time-expansion (slower playback) for $\beta<1$ and time compression (faster playback) for $\beta>1$.  Subjective testing is undertaken in order to justify the quality of the processing.  However the testing is expensive and time consuming.  Recently, an objective measure was developed by \citet{Roberts_2020_OMOQ} that estimates quality, with loss and correlation equivalent to the 97th and 82nd percentiles of subjective sessions in \citet{Roberts_2020_SMOS}.   However, this method requires reference and test signals, and additional interpolation to align low-bandwidth representations of the signals.  In this work, we propose multiple single-ended objective measures of quality for audio processed with TSM.  A convolutional or recurrent neural network front-end generates data-driven features, while a Fully-Connected Neural Network (FCNN) back-end predicts the overall quality. The measures are trained using the dataset of \citet{Roberts_2020_SMOS}, where the dataset is referred to as TSMDB from this point.

Subjective evaluation is the gold standard for evaluating quality of speech and audio processing.  Participants are asked to rate the processing quality of audio files, often using ratings of Bad, Poor, Fair, Good and Excellent that map linearly to the interval [1,5].  Opinion scores are then averaged, giving a Mean Opinion Score (MOS) per file.  This process however is lengthy and expensive.  Consequently, many objective measures of quality have been proposed to predict MOS.

Objective measures can be classified into double-ended (DE) (invasive) and single-ended (SE) (non-invasive) methods. The former calculates differences between reference and processed signal pairs, while the latter operates solely on the processed signal.  This allows non-invasive measures to be used in a variety of use cases such as testing of in-service real-time systems using multiple tests through a signal path, as in \citep{Kim_2005} and \citep{Falk_2006}.  SE measures have seen considerable use for speech quality (\citet{Falk_2006}, \citet{Malfait_2006} and \citep{Gamper_2019}), while audio quality measures such as \cite{Thiede_2000}, \citet{Beerends_2013_POLQA}, \citet{Huber_2006} and \citet{Chinen_2020} are DE.

SE methods are often compared to baseline DE measures such as Perceptual Evaluation of Speech Quality (PESQ) \citep{Rix_2001_PESQ} and Perceptual Evaluation of Audio Quality (PEAQ) \citep{Thiede_2000}.  However, there is no standard for objective quality of TSM, with minimal published literature on the topic.  The total published research is found in the following papers.  \citet{Fierro_2020} published preliminary work towards an objective measure, initial measures were published by \citet{Roberts_2020_SMOS} and formalised in \citet{Roberts_2020_OMOQ}, which  extended PEAQ with additional features and explored synchronization of reference and time-scale processed signals.

The DE method of \citet{Roberts_2020_OMOQ}, referred to as OMOQDE from this point, considered six methods of signal alignment before calculation of PEAQ features --- in addition to hand-crafted features specific to the artefacts of time-scaled signals.  Formulated as a regression problem, an FCNN was used to predict the MOS targets of the TSMDB.  Alignment of reference and processed signals was achieved by interpolating the reference magnitude spectrum to the length of the processed spectrum before feature extraction.  OMOQDE performance was improved by including reference files during training.  Baseline performance was obtained by retraining the PEAQ Basic FCNN to Subjective Mean Opinion Score (SMOS) values.  OMOQDE achieved an average Pearson Correlation Coefficient (PCC) ($\overline{\rho}$) of 0.864 and an average Root Mean Square Error (RMSE) loss ($\overline{\mathcal{L}}$) of 0.490 using the MOS range of 1-5, for the training, validation, and test sets.  OMOQDE was able to resolve statistically significant differences in mean quality between TSM methods of 0.1 MOS.  A distance measure that penalised overfitting was used to select the ideal network, and is discussed further in Section~\ref{sec:Method}.

OMOQDE was trained using the TSMDB, which contains a training subset of 5280 files and a testing subset of 240 files.  Traditional and state-of-the-art TSM methods were used in the training subset, with newer esoteric methods used in the testing subset.  This resulted in no overlap between the TSM methods, time-scale ratios, or reference signals.  The TSM methods included:
\begin{itemize}
    \item Phase Vocoder (PV) \citep{Portnoff_1976} 
    \item Identity Phase-Locking Phase Vocoder (IPL) and Scaled Phase-Locking Phase Vocoder (SPL) \citep{Laroche_Dolson_1999_IPL}
    \item Waveform Similarity Overlap Add (WS) \citep{Verhelst_Roelands_1993}
    \item Fuzzy Epoch Synchronous Overlap-Add (FES) \citep{Roberts_2019_FESOLA}
    \item Harmonic Percussive Separation Time-Scale Modification (HP) \citep{Driedger_Muller_Ewert_2014}
    \item Mel-Scale Sub-band Modelling (uTVS) \citep{Sharma_2017} and the version used in subjective testing ($\overline{\textrm{uTVS}}$)
    \item Elastique (EL) \citep{elastique}
    \item Phase Vocoder using fuzzy classification of bins (FPV) \citep{Damskagg_2017}
    \item Non-Negative Matrix Factorization Time-Scale Modification (NMF) \citep{Roma_Green_Tremblay_2019}
    \item PhaVoRIT ($\overline{\textrm{IPL}}$ and $\overline{\textrm{SPL}}$) \citep{Karrer_Lee_Borchers_2006}
    \item Epoch Synchronous Overlap-Add (ES) \citep{Rudresh_2018}.
\end{itemize}
Quality labels were provided as MOS and median opinion scores, calculated before and after session normalization in the interval [1,5].  The scores were collated from 42,529 ratings by 263 participants in 633 sessions, with a minimum of seven ratings per file.  All files in the dataset have a single channel, a sampling rate of 44.1kHz, and bit depth of 16 bits.  Some reference files are stereo, and were converted to a single channel by summation and normalization to the interval [-1,1].

Deep learning is often used in objective measures of quality.  Convolutional Neural Networks (CNNs) \citet{Lecun_2015} are commonly used on spatial domain tasks, such as image classification.  They have also found use in speech and audio  due to the spatio-temporal representation of short-time frequency analysis, as in \citet{Gamper_2019}.  CNNs learn weights of convolutional kernels which are applied successively creating higher order representations of the signal.  Recurrent Neural Networks (RNNs) differ from standard fully-connected networks through the inclusion of a memory cell and are suited to time-series data.  In this paper, Long Short-Term Memory (LSTM) \citep{Hochreiter_1997_LSTM} and Gated Recurrent Units (GRU) \citep{Cho_2014_GRU} are the used cell types.  LSTM cells are controlled by three gates, input, output and forget, which determine what information is added or removed from the cell.  GRU is a variant of LSTM that removes the output gate and has fewer parameters.  Bidirectional Recurrent Neural Networks \citep{Schuster_1997} extend RNNs with forward and backward passes over the time-series.  

Introduced by \citet{Davis_1980_MFCC}, Mel Frequency Cepstral Coefficients (MFCCs) have found extensive use as a lower bandwidth transformed signal representation in speech processing, as in \citet{Nicolson_2018}.  MFCCs are computed by first estimating the periodogram of the short-time power spectrum. A bank of triangular-shaped filters spaced uniformly on the mel-scale is then applied, resulting in the energy of each filter. The logarithm of the filterbank energies is then taken, followed by a Discrete Cosine Transform to decorrelate the filterbank energies. Differential and acceleration coefficients are often used to give an indication of the dynamics of the MFCCs and are generally known as Deltas ($\textrm{D}$) and Delta-Deltas ($\textrm{D}^\prime$).

The paper is organized as follows: Section~\ref{sec:Method} presents the proposed OMOQSE methods; Section~\ref{sec:Results} presents network results as well as a comparison of TSM algorithms.  Availability, future research and conclusions are presented in Sections~\ref{sec:6}, \ref{sec:7} and \ref{sec:8} respectively.
 
\section{Method}
\label{sec:Method}
First we describe the audio processing.  Signals were prepared by normalizing to the interval [-1,1] and trimming silence at the beginning and end of the signal.  Silence was determined, according to \citet{ITU_BS1387_PEAQ}, as the first and last time the sum of four consecutive samples is greater than 0.0061.  The magnitude spectrum ($|X|$), magnitude and phase spectra ([$|X|$;$\angle X$]), power spectrum ($|X|^2$), MFCCs, MFCCs and $\textrm{D}$ ([MFCCs;$\textrm{D}$]), and MFCCs, $\textrm{D}$ and $\textrm{D}^\prime$ ([MFCCs;$\textrm{D}$;$\textrm{D}^\prime$]), where [$\cdot$ ; $\cdot$] is concatenation, were tested during development.  The magnitude, phase, and power spectra used a frame length of $N=2048$ samples, an overlap of N/2 and a Hann window.  MFCCs were of length 128, with $\textrm{D}$ and $\textrm{D}^\prime$ width nine from $t-4$ to $t+4$ with respect to the current time-step.  Overall or per frequency-bin standardization of the input features was explored.

Due to the variable length of the input signal, truncating and duplicating the signal were explored.  For the CNN, sequences were truncated to the overall minimum length (L), starting from a different random location in each epoch.  During testing, the OMOS was averaged over 16 segments to capture more information of the processed signal, for a wider selection of input signals.  Repeating the input signal to the duration of the longest signal was also considered for GRU-FT, however as LSTM and GRU operate sequentially on each frame, input signals were used in their entirety.

Prior to network training, target scores were scaled to the interval [0,1] using
\begin{equation}
\label{eq:Norm3}
    SMOS = \frac{SMOS-1}{4} \quad .
\end{equation}

\begin{figure*}[ht]
    \centering
    \includegraphics[trim= {0 0 0 0},clip,height=2in]{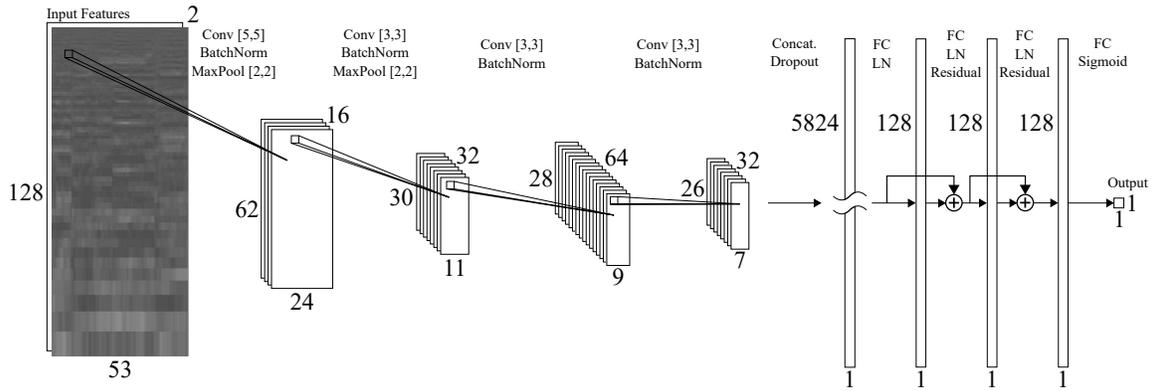}
    \caption{Proposed CNN dataflow. Kernel sizes in brackets, numbers denote layer size and number of channels, FC is a fully-connected layer, LN is layer normalization, ReLU activation used unless specified.}
    \label{fig:CNN}
\end{figure*}

The proposed CNN data structure is shown in Fig.~\ref{fig:CNN}.  It contains four convolution layers, of filter sizes 16, 32, 64 and 32, with batch normalization and a 5x5 kernel for the first layer, and 3x3 for the remaining layers.  The first two convolutional layers are followed by max pooling layers, with 2x2 kernels and 2x2 stride.  After concatenation and 10\% dropout, three fully-connected layers of output size 128 are used. The final layer has a single output.  Rectified linear unit (ReLU) activation is used throughout, except for the output layer, where the Sigmoid activation is used. Residual connections around the second and third fully-connected layers are used.  Root Mean Square Error (RMSE) is used as the loss function.  Features were concatenated in time-aligned input panes.

\begin{figure*}[ht]
    \centering
    \includegraphics[trim= {0 0 0 0},clip,height=2in]{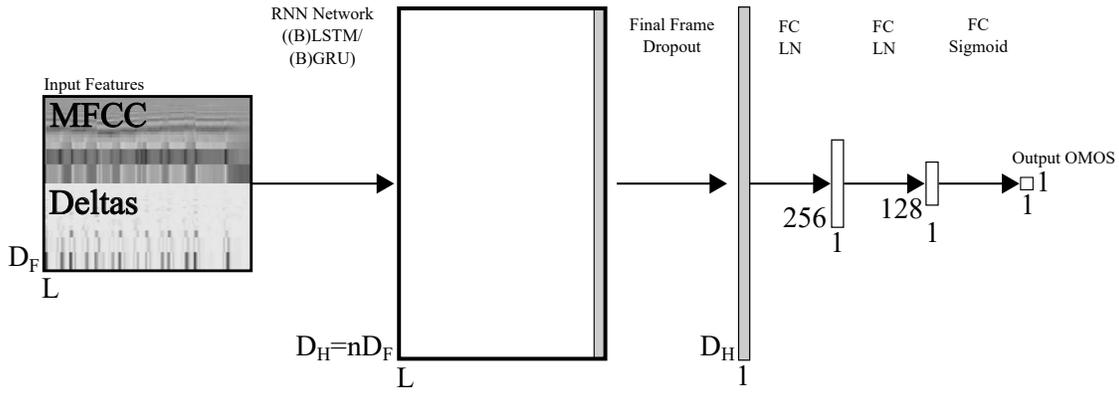}
    \caption{Proposed RNN FF dataflow. $D_F$ is feature depth, $D_H$ is Hidden Dimensions, n is the number of directions, numbers denote layer sizes, FC is a fully-connected layer, LN is layer normalization, ReLU activation used unless specified.}
    \label{fig:RNN}
\end{figure*}

The proposed final-frame (FF) model for LSTM, BLSTM, GRU and BGRU networks can be seen in Fig.~\ref{fig:RNN}.  FF RNN models use backpropagation through time to learn from the error between the final output and the SMOS. The total feature dimension ($D_F$) is set by the concatenation of input features.  For the proposed network using [MFCCs;$\textrm{D}$] features, $D_F$ is 256.  Two RNN layers were used with the memory layer size ($D_H$) set to the number of directions ($n$) multiplied by $D_F$. L is the sequence length and ranged from 53 to 2179 frames.  An RNN architecture of many-to-one was used, with the final frame used as input to an FCNN after 10\% dropout.  The FCNN contained three layers of output size 256, 128 and 1, respectively.  Layer normalization and ReLU activation were used for layers 1 and 2, while Sigmoid activation was used for the output layer.  Again, RMSE is used as the loss function.  Magnitude, Phase and Power spectra ($D_F=1025, D_H=512$) were also explored as input to this network.

\begin{figure*}[ht]
    \centering
    \includegraphics[trim= {0 0 0 0},clip,height=2in]{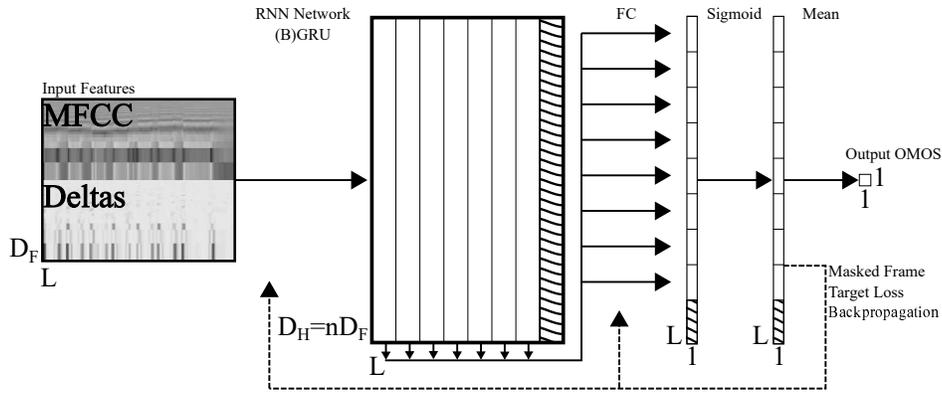}
    \caption{Proposed GRU FT network dataflow. $D_F$ is feature depth, $D_H$ is hidden dimensions, n is the number of directions, L is sequence length, numbers denote layer sizes, FC is a fully-connected layer and hashed sections are zero-padding to longest file in mini-batch.}
    \label{fig:GRU}
\end{figure*}

The proposed frame-target (FT) model for GRU and BGRU networks (GRU-FT and BGRU-FT) can be seen in Fig.~\ref{fig:GRU}. Two GRU layers of $D_H=256$ with 10\% dropout were used in a similar structure to the previous RNN.  However, a single fully-connected layer with sigmoid activation reduces feature dimensionality to $L\times1$.  The Mean Square Error (MSE) between the target SMOS and each frame estimate is used as loss.  Frame targets are averaged for the length of the sequence to calculate the OMOS.  As this calculation is independent of training, median, minimum and maximum values of frame targets were also considered.  Minimum frame targets were considered as quality evaluation of time-scaled signals is a degradation style analysis, where subjective quality is heavily influenced by the quality of the worst part of the signal.  Due to the inverse correlation between time-scale ratio and SMOS for signals that have been slowed down, an inverse exponential relationship between the number of frames at the time-scale and the time-scale itself, possibly leading to difficulty in estimating quality for signals that have been sped up.  The impact of this was explored by training on signals truncated to the minimum signal length and on signals repeated to the maximum signal length.

10\% of the training dataset was reserved for validation.  The CNN was trained for 100 epochs using a mini-batch size of 132, while RNNs were trained for 30 to 60 epochs with a mini-batch size of 48.  A learning rate of $1e^{-4}$ was used in most cases, with $1e^{-5}$ if network performance stopped improving within the first 10 epochs.  AdamW \citep{AdamW_2017} was used as the optimizer for all networks.  Loss for backpropagation was calculated using estimates in the interval of [0,1]. Reported loss values ($\mathcal{L}$) were calculated using RMSE and estimates scaled back to the original interval of [1,5], for comparison with OMOQDE.  As the prediction of opinion scores for novel TSM methods is the use case, early stopping based on validation loss was not used.  The optimal epoch minimised the distance measure of \citet{Roberts_2020_OMOQ}, where the minimum overall distance ($\mathcal{D}$), is calculated by
\begin{equation}
\label{eq:Overall_Distance1}
    \mathcal{D} = \|[\hat{\rho}, \hat{\mathcal{L}}]\|_{_2} \quad ,
\end{equation}
where $\hat{\rho}$ and $\hat{\mathcal{L}}$ are calculated by
\begin{equation}
\label{eq:Overall_Distance2}
    \hat{\rho} = \|[1-\overline{\bm{\rho}},\Delta\bm{\rho}]\|_{_2} \quad ,
\end{equation}
\begin{equation}
\label{eq:Overall_Distance3}
    \hat{\mathcal{L}} = \|[\overline{\bm{\mathcal{L}}},\Delta\bm{\mathcal{L}}]\|_{_2} \quad .
\end{equation}
where $\bm{\rho}=[\rho_{tr}, \rho_{val}, \rho_{te}]$, 
$\bm{\mathcal{L}}=[\mathcal{L}_{tr}, \mathcal{L}_{val}, \mathcal{L}_{te}]$, 
$tr$, $val$ and $te$ denote training, validation and testing,
$\overline{\bm{\mathcal{L}}}$ is the mean of $\bm{\mathcal{L}}$, 
$\overline{\bm{\rho}}$ is the mean of $\bm{\rho}$, 
$\Delta\bm{\rho} = max(\bm{\rho})-min(\bm{\rho})$ and 
$\Delta\bm{\mathcal{L}} = max(\bm{\mathcal{L}})-min(\bm{\mathcal{L}})$.  This scheme limits over-training and allows for the novel artefacts of the test subset to inform the chosen optimal network, without their direct use in training.

An evaluation set of 6000 files, published as part of \citet{Roberts_2020_OMOQ}, was generated from the reference files in the test set.  20 new time-scales in the range of $0.22 < \beta < 2.2$, with all TSM methods listed in Section~\ref{sec:1} used to process the reference files.  During evaluation, averages do not include $\beta=0.2257$ as the minimum for EL is $\beta=0.25$, or $\beta=1$ as it should result in a unity system.  This is not always the case, and can be useful for determining method performance, but it has been excluded from the analysis.

\section{Results}
\label{sec:Results}
\subsection{Network Performance}
\label{subsec:Results:2}
\begin{figure}[ht]
    \centering
    \includegraphics[trim= {0 0 0 0},clip,width=\myscale\linewidth]{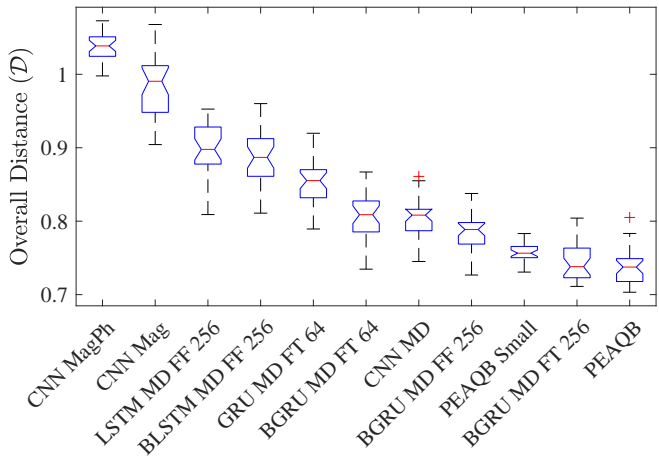}
    \caption{Box plot of best distance measure for 30 seeds of each network configuration, ordered by median $\mathcal{D}$. $|X|$ is denoted by Mag, $\angle X$ by Ph, [MFCCs;$\textrm{D}$] by MD and hidden size denoted by 64 or 256.}
    \label{fig:Overall_Distance_Boxplot}
\end{figure}

A wide range of testing and network configurations were considered during the development of the proposed measures.  Network hyper-parameters were optimized through a systematic non-exhaustive search.  Deterministic training of all networks was conducted using seeds from 0 to 29.  Figure~\ref{fig:Overall_Distance_Boxplot} shows the box plot distribution of the best $\mathcal{D}$ for each seed, where lower is better.

Median overall distance ($\widetilde{\mathcal{D}}$) and the best case $\bm{\mathcal{D}}$ with associated $\overline{\bm{\mathcal{L}}}$, $\Delta\bm{\mathcal{L}}$, $\overline{\bm{\rho}}$ and $\Delta\bm{\rho}$ values can be found in Table~\ref{tab:RMSE_PCC}, along with ($\mathcal{L}_{te}$) and ($\rho_{te}$).  While the improvement in performance appears linear in Fig.~\ref{fig:Overall_Distance_Boxplot}, many network configurations have not been included.  Most networks trained with [MFCCs;$\textrm{D}$] achieved $0.55<\mathcal{L}_{te}<0.67$, with only BGRU-FT achieving $\mathcal{L}_{te}>0.68$ or $\mathcal{D}<0.72$.  This appears to be the $\mathcal{L}_{te}$ and $\mathcal{D}$ limit for these network configurations and input features, even with $\rho_{tr}$ approaching 1 when allowed to over-train. 

\input{./Table1} 

The results in Table~\ref{tab:RMSE_PCC} can be summarised as follows. The proposed CNN achieved a best $\overline{\bm{\mathcal{L}}}$ of 0.608, $\overline{\bm{\rho}}$ of 0.771, $\mathcal{L}_{te}$ of 0.801 and $\rho_{te}$ of 0.637 placing it at the 74th and 32nd percentiles of subjective sessions for $\overline{\bm{\mathcal{L}}}$ and $\overline{\bm{\rho}}$ respectively.  The proposed BGRU-FT network achieved a best $\overline{\bm{\mathcal{L}}}$ of 0.576, $\overline{\bm{\rho}}$ of 0.794, $\mathcal{L}_{te}$ of 0.762 and $\rho_{te}$ of 0.682 placing it at the 84th and 39th percentiles of subjective sessions for $\overline{\bm{\mathcal{L}}}$ and $\overline{\bm{\rho}}$ respectively. 


To given an indication of what the networks may be learning, correlation between OMOQSE OMOS and OMOQDE features was calculated for CNN and BGRU-FT networks.  No significant correlation was found with maximum correlations of 0.210 and 0.206 for CNN and BGRU-FT respectively.

Several trends were seen across testing.  Networks trained using [MFCCs;$\textrm{D}$] features out-performed those trained using [MFCCs;$\textrm{D}$;$\textrm{D}^\prime$], as well as solely MFCCs, magnitude spectra, magnitude and phase spectra, and the power spectrum. In all cases, magnitude only features out-performed combined magnitude and phase features.  Decreased performance due to the inclusion of phase is likely due to its noise-like quality.  Results for networks trained on the power spectrum are not shown in plots to increase comprehension.  Improved performance was found using MFCCs generated with Librosa over TorchAudio, with identical settings.  For RNNs, FT measures outperformed FF measures, GRU outperformed LSTM, and bidirectional networks generally outperformed single direction networks for the same input features and network size.  As such, RNN analysis will focus on BGRU-FT, alongside analysis of CNN performance.  

\begin{figure}[ht]
    \centering
    \includegraphics[trim= {0 0 0 0},clip,width=\myscale\linewidth]{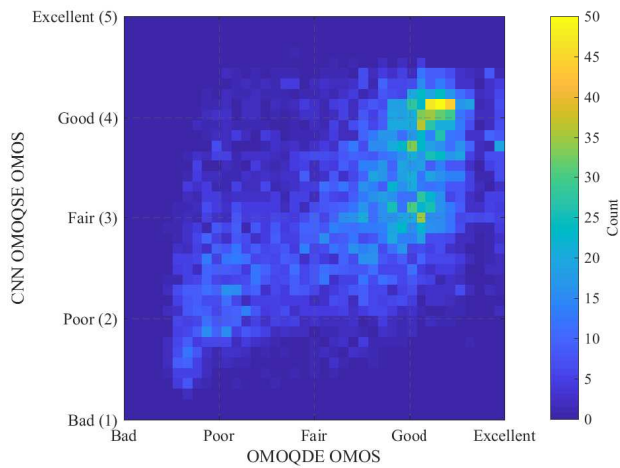}
    \caption{[Color Online] OMOS confusion matrix for CNN OMOQSE and OMOQDE.}
    \label{fig:OMOQ_vs_CNN_OMOQSE}
\end{figure}

The CNN improved significantly through the use of [MFCCs;$\textrm{D}$] over $|X|$, $\angle X$ and $|X|^2$, with similar performance to FF RNNs.  Normalization of input spectra reduced network performance.  Small gains were found through optimising the kernel size, however the maximum kernel size was limited by the length of the shortest file.  Repeating signals to the length of the longest files decreased network performance, as did using a combination of repeating or truncating to 500 or 1000 frames.  The CNN predicts across most of the OMOS range, shown in Fig.~\ref{fig:OMOQ_vs_CNN_OMOQSE}, and achieves a correlation of 0.564872 with the OMOQDE OMOS.  Loss and correlation can be found in Table~\ref{tab:RMSE_PCC}.
%

\begin{figure}[ht]
    \centering
    \includegraphics[trim= {0 0 0 0},clip,width=\myscale\linewidth]{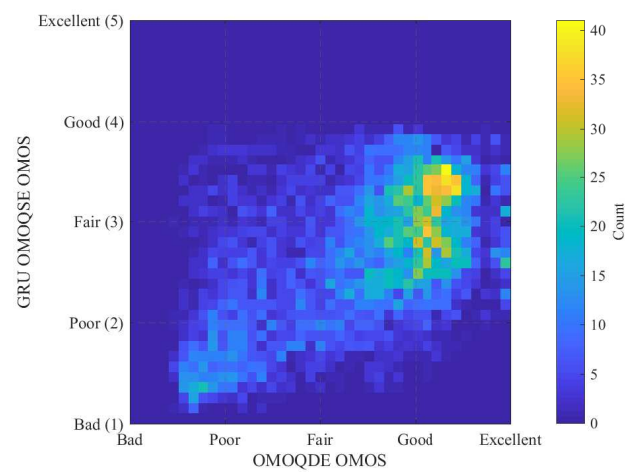}
    \caption{[Color Online] OMOS confusion matrix for BGRU-FT OMOQSE and OMOQDE.}
    \label{fig:OMOQ_vs_GRU_OMOQSE}
\end{figure}

The BGRU-FT was found to give the best performance of the tested SE networks according the distance measure, and gives similar performance to OMOQDE trained using PEAQ Basic features.  $\overline{\bm{\mathcal{L}}}$ and $\overline{\bm{\rho}}$ are improved over PEAQB networks, despite worse $\mathcal{L}_{te}$ and $\rho_{te}$, resulting in larger $\Delta\bm{\mathcal{L}}$ and $\Delta\bm{\rho}$ values, shown in Table~\ref{tab:RMSE_PCC}.  When collapsing estimated frame targets, no significant difference was found between mean or median of predictions, while selecting the minimum or maximum prediction reduced performance.  A short-coming of most BGRU-FT networks trained is the lack of predictions for $OMOS>4$, which can be seen in Fig.~\ref{fig:OMOQ_vs_GRU_OMOQSE}, where the correlation is 0.549.  A hidden size of 256 outperformed 64, 128 and 512 sizes, with 10\% dropout outperforming 0\%, 25\% and 50\%.  Including $\textrm{D}^\prime$ was found to reduce performance, as did increasing the number of MFCCs to 256.  Multiple fully-connected layers were also explored, but did not improve performance.  FT RNNs slightly improved performance over FF RNNs, with FF improvements following BGRU-FT results, with the best FF network shown in Fig.~\ref{fig:Overall_Distance_Boxplot} and Table~\ref{tab:RMSE_PCC}.

\begin{figure}[ht]
    \centering
    \includegraphics[trim= {0 0 0 0},clip,width=\myscale\linewidth]{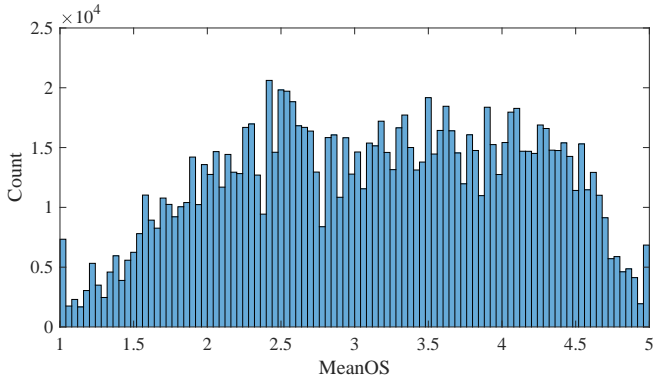}
    \caption{Distribution of frames per MOS in training set.}
    \label{fig:Frame_MOS_Distribution}
\end{figure}

Experiments showed that using truncated random segments with BGRU-FT reduced performance, as did extending signals through repetition.  Repeating input for the CNN also reduced performance.  While the number of frames for $\beta<<1$ in the training subset is significantly greater than for $\beta>>1$, the number frames is relatively uniform for $2\lesssim SMOS \lesssim 4.5$, see Fig.~\ref{fig:Frame_MOS_Distribution}.  The reduced number of frames for $1\leq SMOS \lesssim 1.5$ and $4.5\lesssim SMOS\leq 5$ may also impact the estimation at outermost OMOS, as seen in Fig.~\ref{fig:OMOQ_vs_GRU_OMOQSE}.  Surprisingly, although truncated segments are used as input to the CNN, the estimates show a wide spread in Fig.~\ref{fig:OMOQ_vs_CNN_OMOQSE}.

\subsection{TSM Algorithm Evaluation}
\label{subsec:Results:3}
\begin{figure*}
\centering
\begin{tabular}{cc}
    \includegraphics[width=0.45\linewidth]{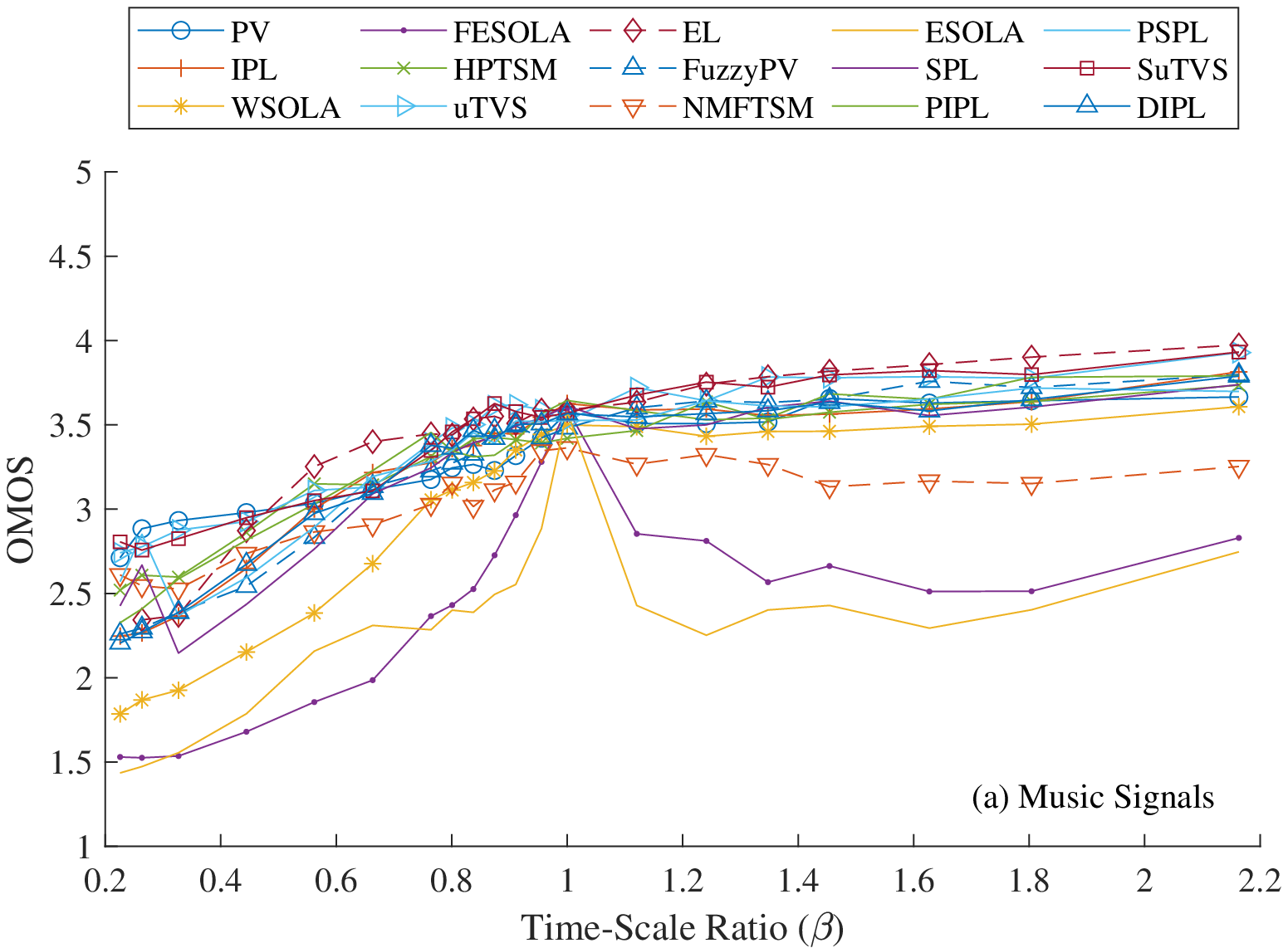} &
    \includegraphics[width=0.45\linewidth]{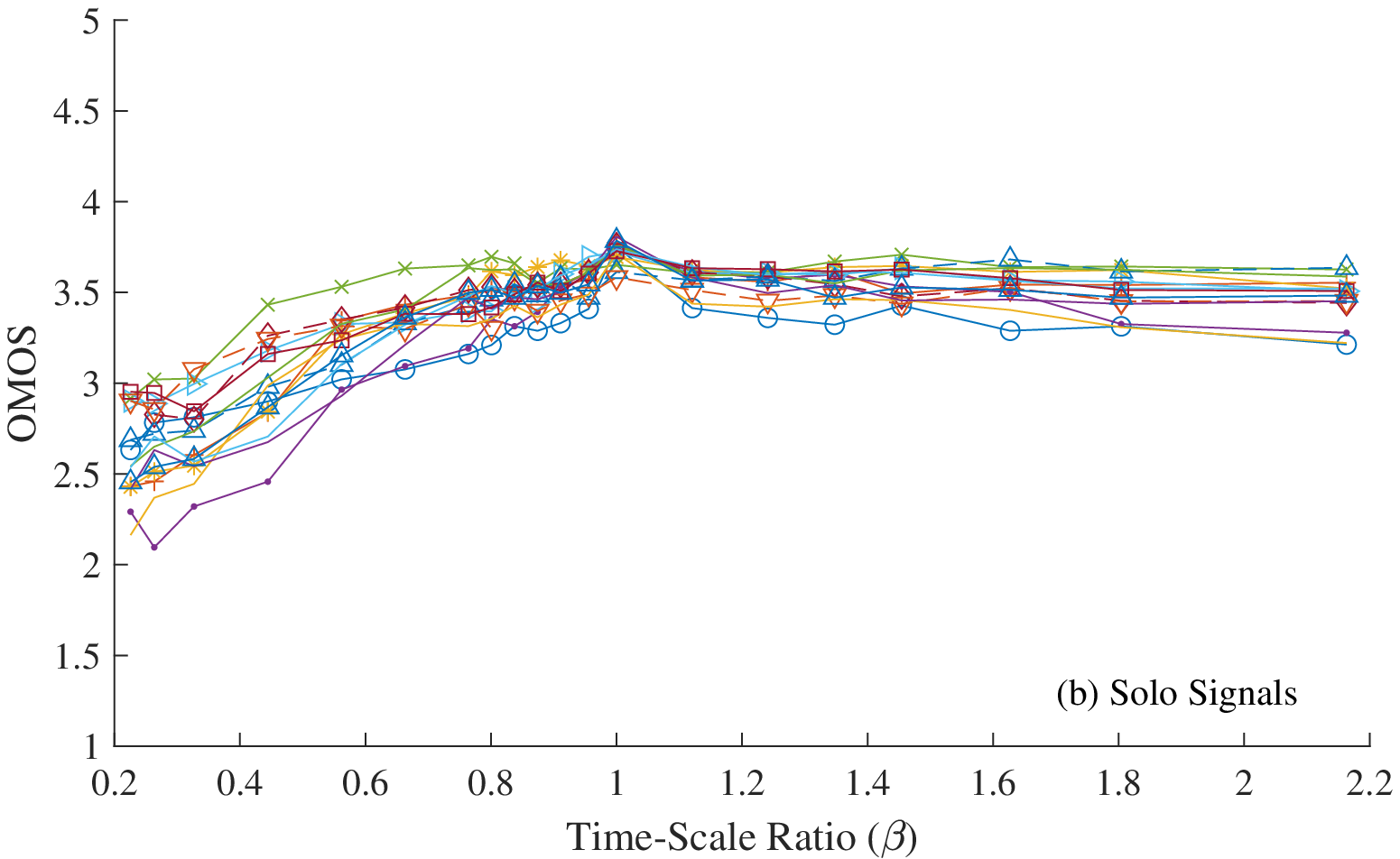} \\
    \includegraphics[width=0.45\linewidth]{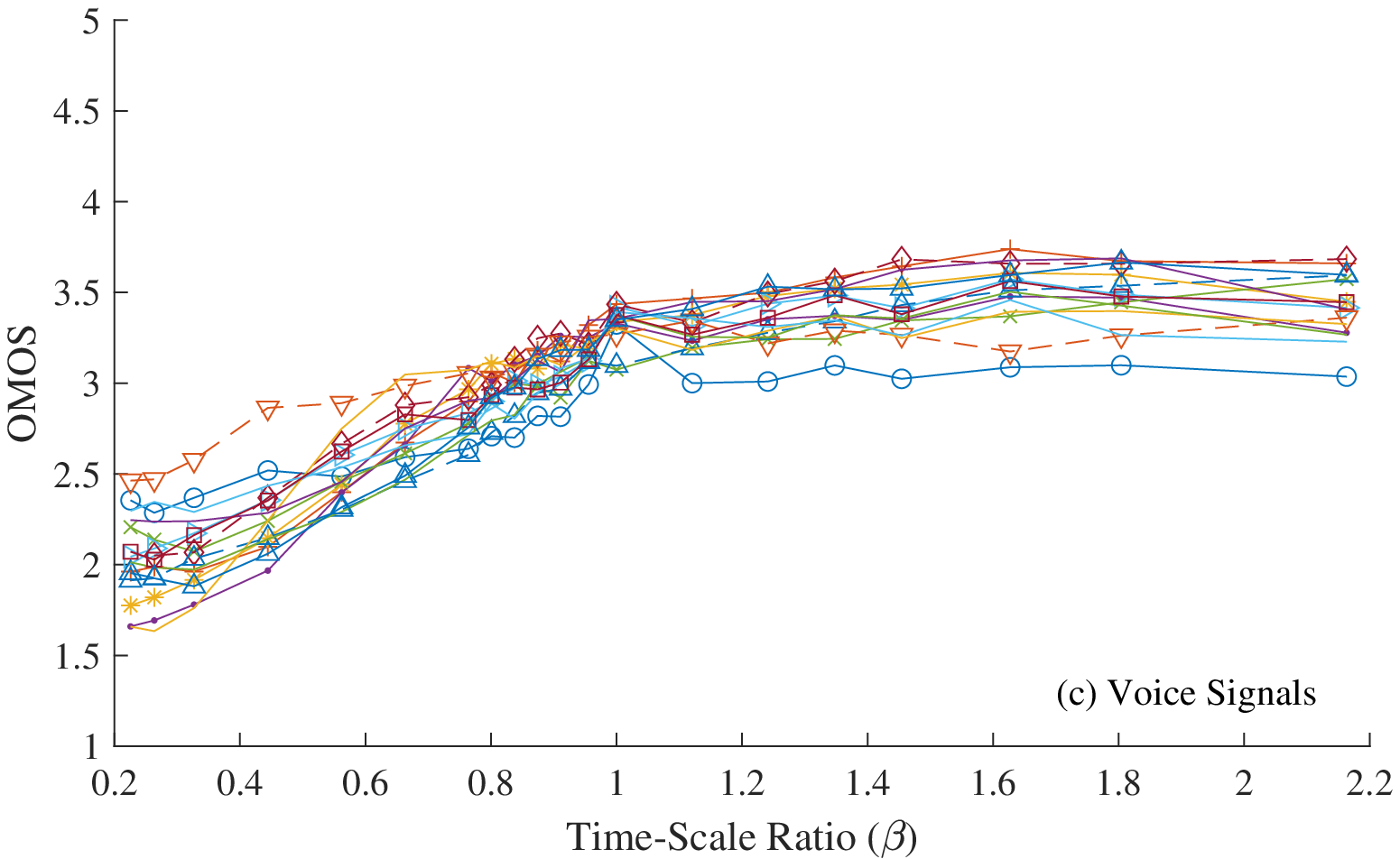} &
    \includegraphics[width=0.45\linewidth]{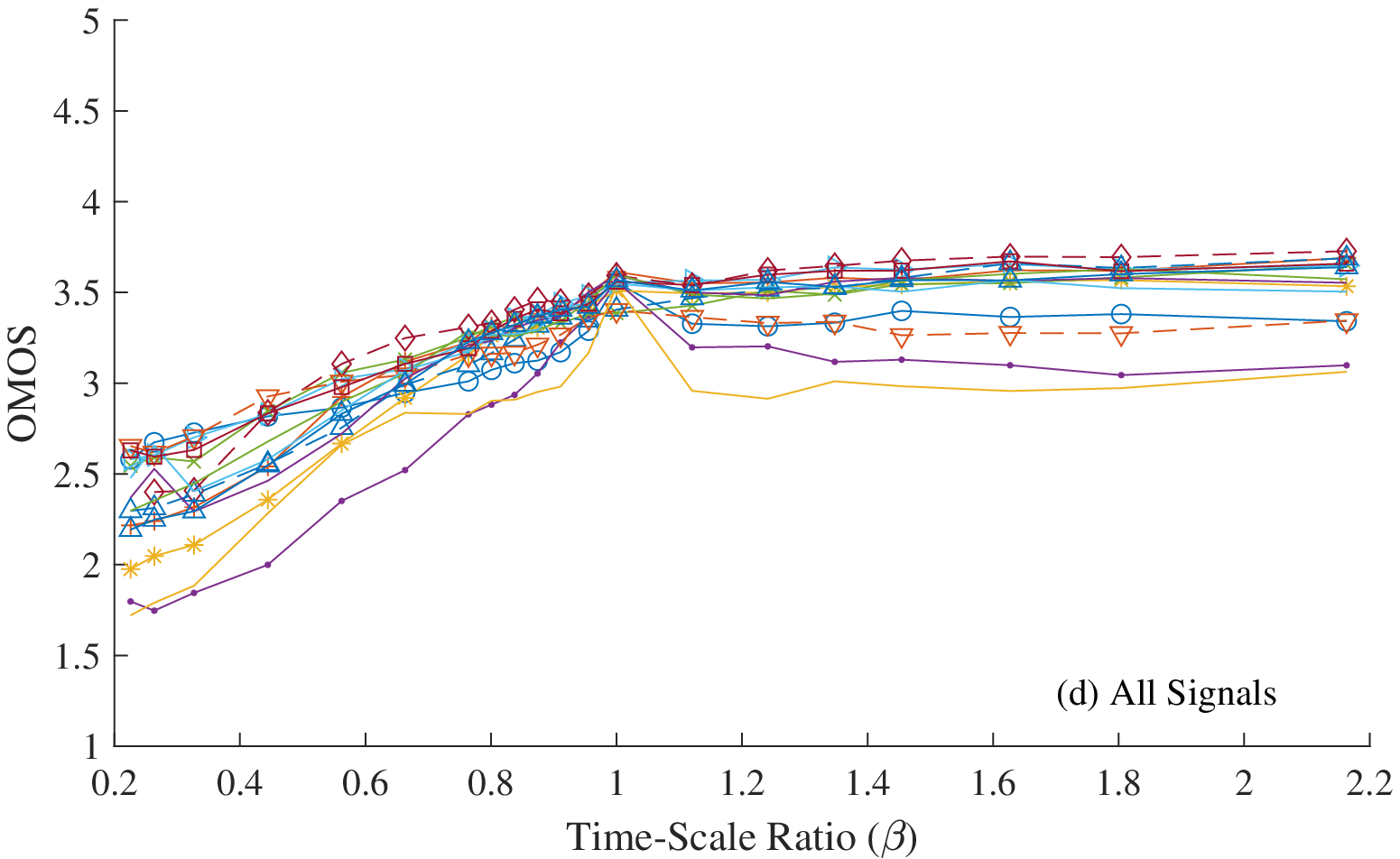}
\end{tabular}
\caption{\label{fig:CNN_Line_Graphs} [Color Online] CNN estimated Mean OMOS for each TSM method as a function of $\beta$ for: (a) Musical signals, (b) Solo signals, (c) Voice signals and (d) All signals combined.}
\end{figure*}

\begin{figure}
    \centering
    \includegraphics[trim= {0 0 0 0},clip,width=\myscale\linewidth]{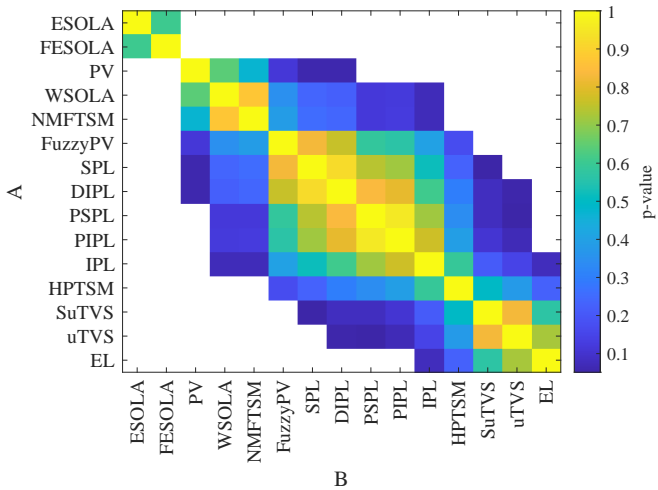}
    \caption{[Color Online] Masked two-sample t-test for all CNN OMOS estimates for each TSM method. Showing $p>0.05$ for TSM method comparisons where the difference in mean is not statistically significant. Unequal means indicated by white.}
    \label{fig:Masked_P_CNN}
\end{figure}

\begin{figure*}
\begin{tabular}{cc}
    \includegraphics[width=0.45\linewidth]{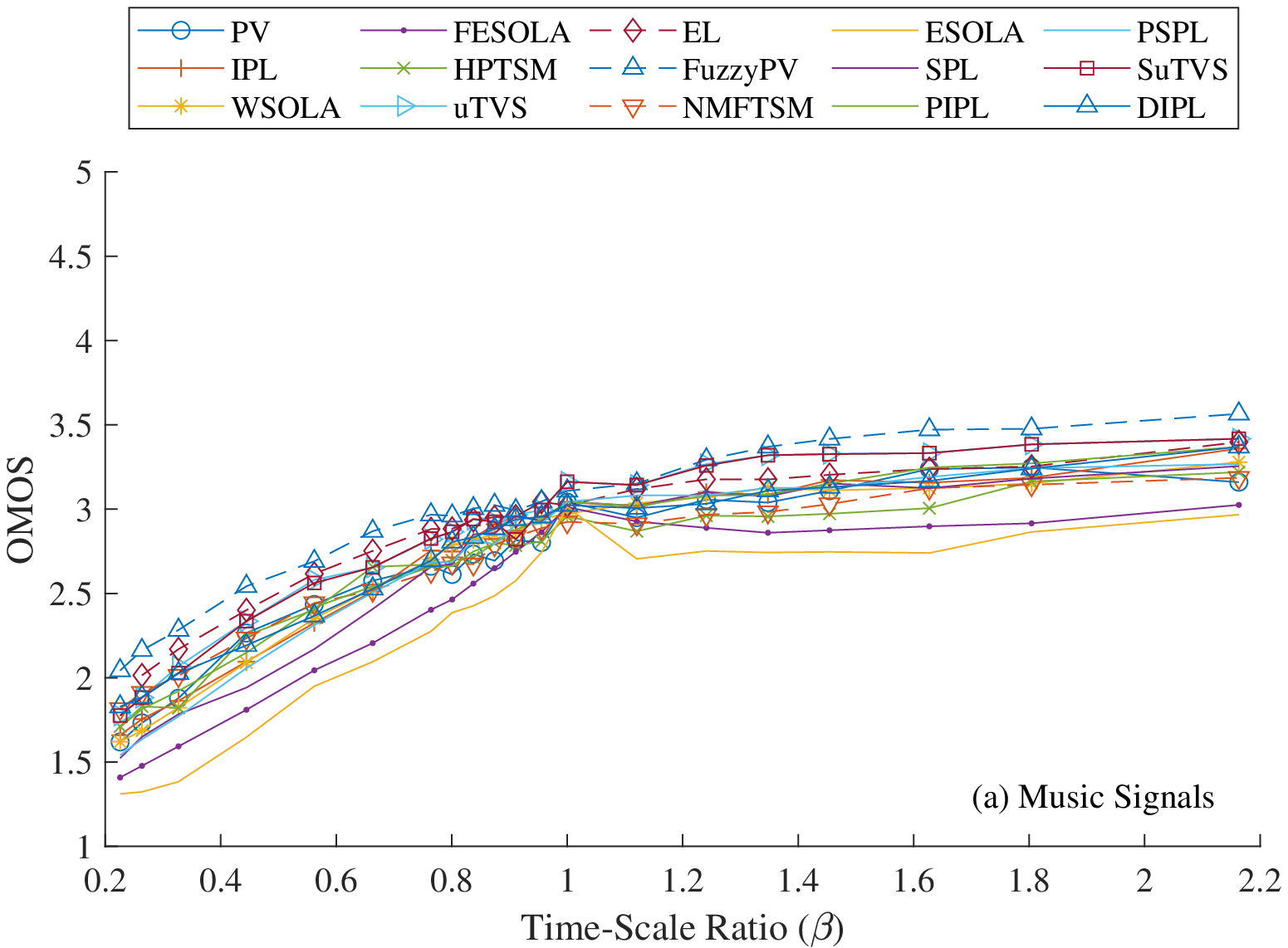} &
    \includegraphics[width=0.45\linewidth]{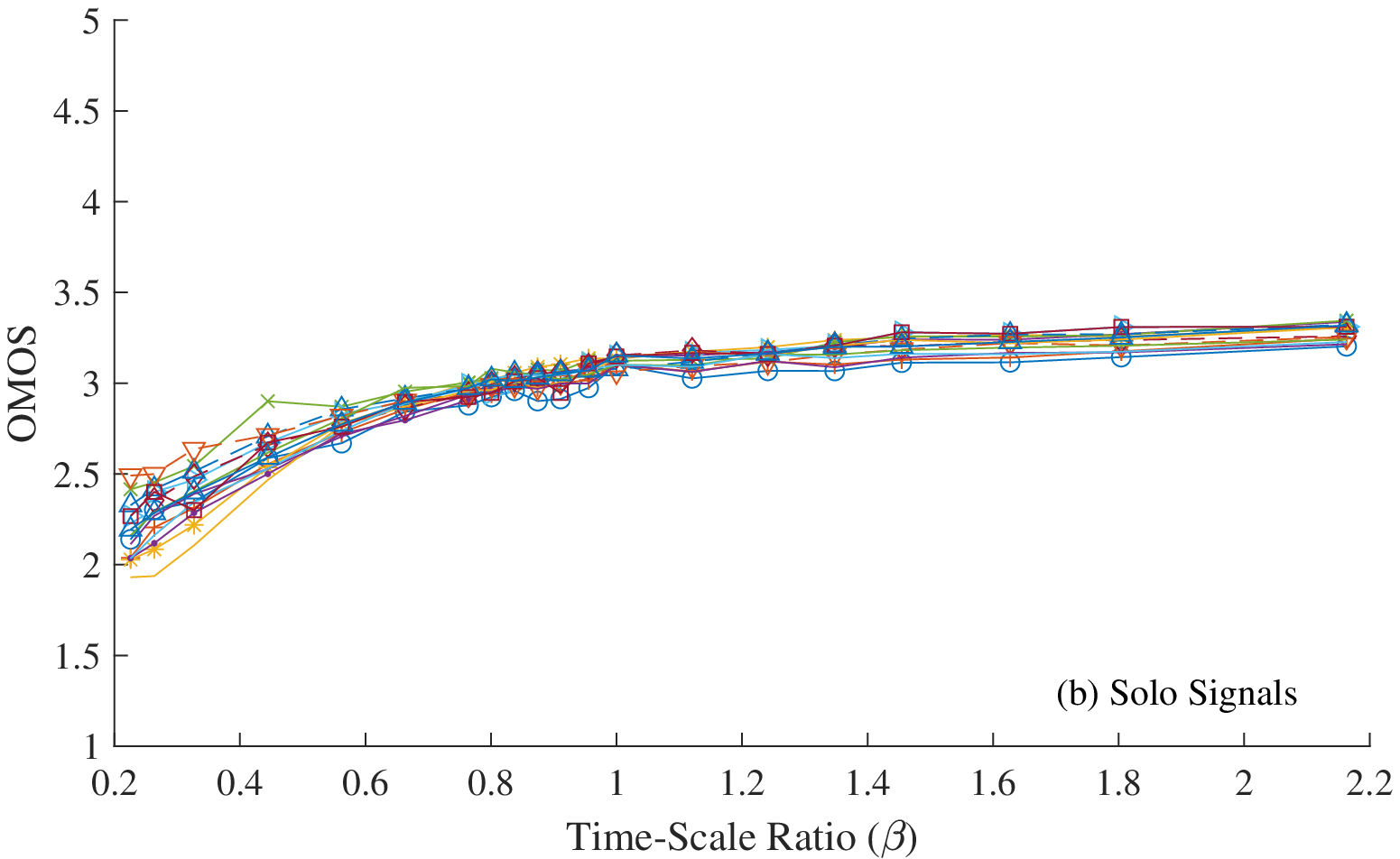} \\
    \includegraphics[width=0.45\linewidth]{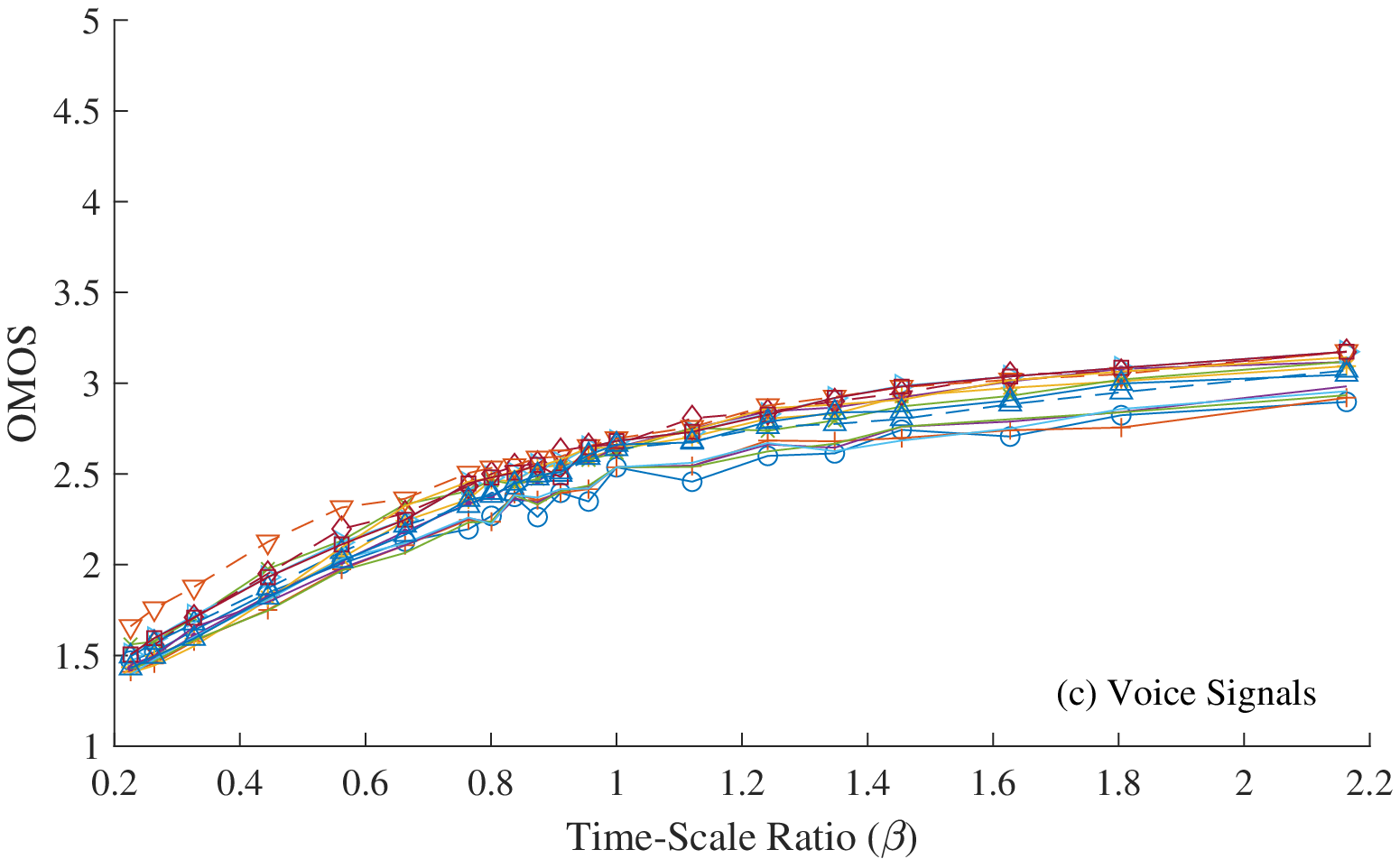} &
    \includegraphics[width=0.45\linewidth]{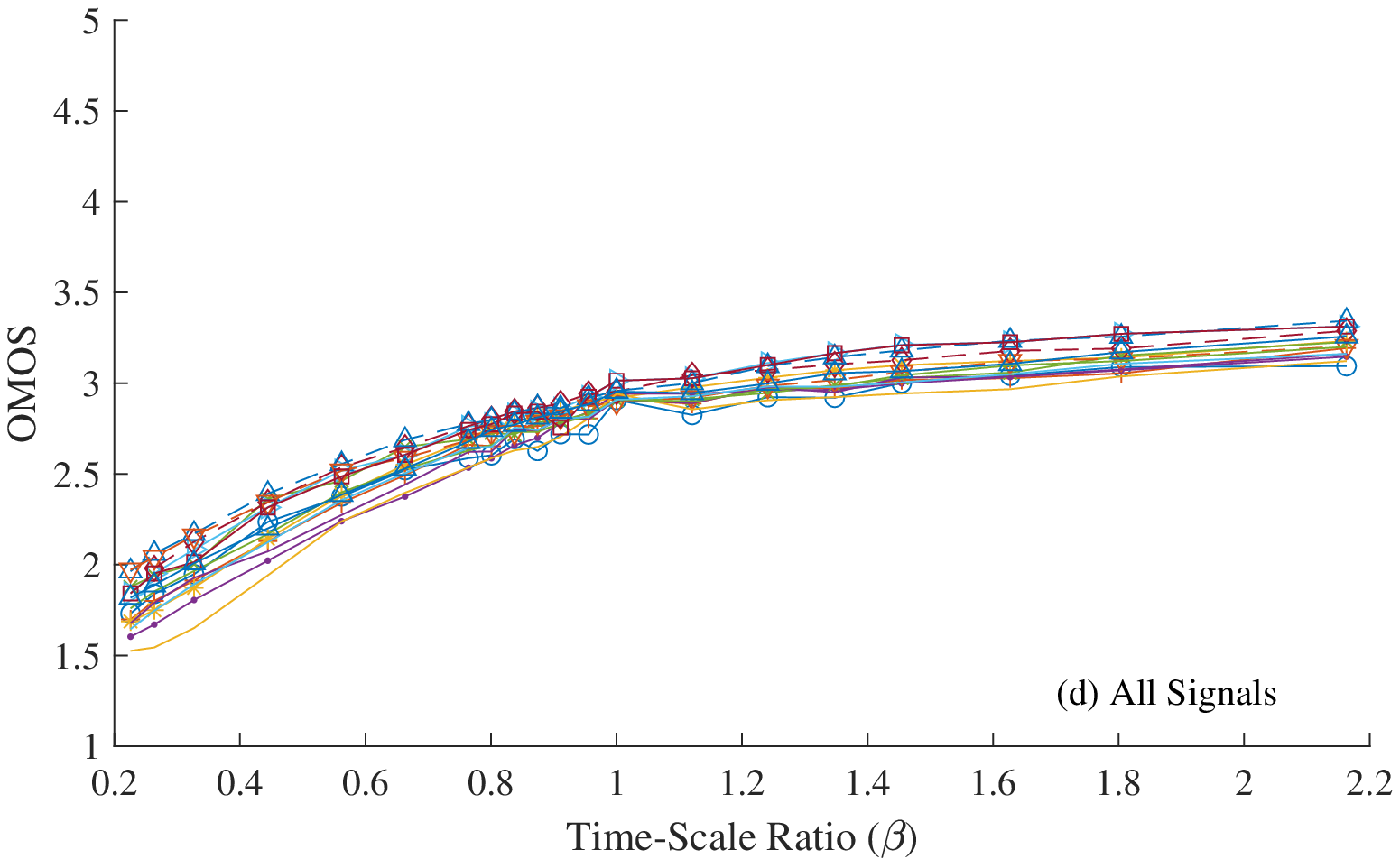}
\end{tabular}
\caption{ \label{fig:BGRU_Line_Graphs} [Color Online] BGRU-FT estimated Mean OMOS for each TSM method as a function of $\beta$ for: (a) Musical signals, (b) Solo signals, (c) Voice signals and (d) All signals combined.}
\end{figure*}

\begin{figure}
    \centering
    \includegraphics[trim= {0 0 0 0},clip,width=\myscale\linewidth]{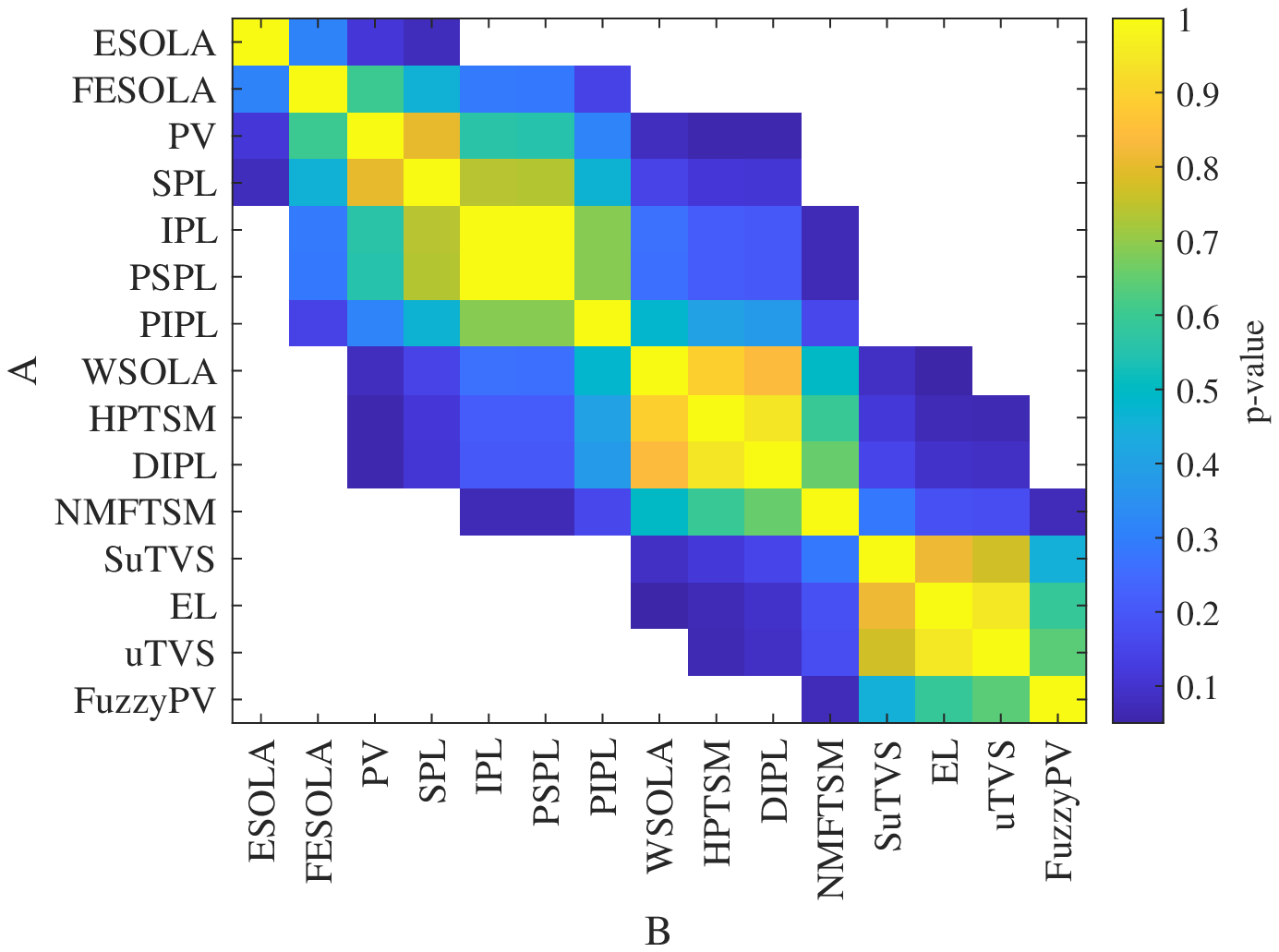}
    \caption{[Color Online] Masked two-sample t-test for all BGRU-FT OMOS estimates for each TSM method. Showing $p>0.05$ for TSM method comparisons where the difference in mean is not statistically significant. Unequal means indicated by white.}
    \label{fig:Masked_P_GRU}
\end{figure}

In this section, TSM methods are evaluated using the aforementioned evaluation set.  Tables ~\ref{tab:CNN_Eval} and~\ref{tab:BGRU_Eval} show average OMOS for each signal class per TSM method ordered by overall mean OMOS, Fig.~\ref{fig:CNN_Line_Graphs} and~\ref{fig:BGRU_Line_Graphs} show average OMOS per TSM method and $\beta$ and Fig.~\ref{fig:Masked_P_CNN} and~\ref{fig:Masked_P_GRU} show TSM methods for which differences in mean are statistically significant.  As in \citet{Roberts_2020_OMOQ}, all results for $\beta=1$ and $\beta<0.25$ were excluded from averaging calculations forming Tables~\ref{tab:CNN_Eval} and \ref{tab:BGRU_Eval}, as time-scaling is applied at $\beta\neq 1$, and the minimum $\beta$ available for EL is $0.25$.  Common trends are presented, followed by CNN and then BGRU-FT analysis.  

Estimation of signals time-scaled using NMF was particularly challenging for all networks.  This is likely due to novel artefacts described by \citet{Roma_Green_Tremblay_2019} and the SMOS distribution skewed towards low scores.  This provides a challenge for network design as novel TSM methods may not have similar artefacts or SMOS distributions to those in the training set. However, the relative rating of EL and FPV to other TSM methods follows that of subjective testing.  As suggested by network $\mathcal{L}_{te}$ and $\rho_{te}$, only a general sense of TSM quality is obtained.  Small details, such as the reduced quality of uTVS used in subjective testing at $\beta\approx1$, are not visible.  The networks have also not learnt the non-linearity of SMOS as a function of $\beta$, continuing to increase for $\beta>1$, seen in Fig.~\ref{fig:CNN_Line_Graphs} and~\ref{fig:BGRU_Line_Graphs}.  The uniform quality of methods at $\beta=1$ is however visible, as is the reduction in TSM quality for $\beta<1$.

For musical files, Fig.~\ref{fig:CNN_Line_Graphs}(a), the CNN differentiates between frequency and time-domain methods, where quality rapidly falls for time-domain methods when $\beta<1$.  WS fairs the best of the time-domain methods, diverging from frequency domain methods for $\beta<0.8$.  The relative improvement in PV quality is also visible for $\beta<0.5$, and the slower falloff of EL.  When averaged, the CNN rates uTVS and subjective uTVS highest followed by EL.  For solo files, (Fig.~\ref{fig:CNN_Line_Graphs}(b), HP exceeds other methods for $\beta<1$.  This class is the only evaluation where the highest OMOS at $\beta=1$.  HP has the highest mean OMOS, followed by Phavorit IPL, both uTVS methods, EL, and WS, as shown in Table~\ref{tab:CNN_Eval}.  Voice file OMOS, Fig.~\ref{fig:CNN_Line_Graphs}(c), is comparatively low for time-domain methods, which is unexpected, as speech is often scaled well by time-domain methods.  The high quality of NMF is also unexpected based on subjective results in \citet{Roberts_2020_SMOS}.  EL has the highest mean OMOS, followed by NMF, SPL and IPL.  All other methods gave similar averaged quality.  


\input{./Table2.tex}

For all CNN OMOS, EL has the highest average rating followed by both uTVS methods and HP.  The overall means can be seen in Fig.~\ref{fig:CNN_Line_Graphs}(d).  The best five methods are separated by $<0.1$ OMOS with a maximum difference of 0.554 for all methods.  The overall low quality of FPV is unexpected given that it builds on IPL, however further analysis is required to determine if the difference statistically significant.  A two-sample t-test ($\alpha=0.05$) of all OMOS shows the null hypothesis, TSM methods having equal means, to be rejected in almost all cases when the absolute difference of mean OMOS is greater than 0.098.  Figure~\ref{fig:Masked_P_CNN} shows p-values for the t-tests that were unable to reject equal means.

BGRU-FT OMOS shows the most variance for music files, Fig.~\ref{fig:BGRU_Line_Graphs}(a).  Again, time-domain methods rate lower.  FPV is rated highest, followed by uTVS and EL.  For multiple TSM methods, OMOS continues to increase for $\beta>1$.  By combining this information with the improvement when $\textrm{D}$ is included as an input feature, we theorise that BGRU-FT is learning a relationship between SMOS and the velocity and duration between $\textrm{D}$ events.  As $\beta$ increases, the time between sound events decreases and the attack portion of the energy envelope becomes sharper.  For solo files, Fig.~\ref{fig:BGRU_Line_Graphs}(b), there is very little variance between methods, with a maximum difference $\approx$0.5 OMOS for $\beta=0.2257$.  Solo files have the highest overall OMOS of the three classes, which is consistent with subjective findings.  HP has the highest mean OMOS, followed by uTVS and FPV.  Voice file OMOS, Fig.~\ref{fig:BGRU_Line_Graphs}(c), shows the lowest TSM quality of the three classes with a continued increase in OMOS for $\beta>1$ across all TSM methods. NMF has the highest mean OMOS, which is unexpected based on \citet{Roberts_2020_SMOS}.  EL is next highest followed by uTVS methods and ES.  The high quality of ES is expected as the method was designed for TSM of speech.  

\input{./Table3.tex}

For overall OMOS, Fig.~\ref{fig:CNN_Line_Graphs}(d), FPV has the highest average rating followed by uTVS methods and EL.  The ordered ranking of methods is close to expected, with only NMF ranking unexpectedly.  This is possibly due to the method retaining the shape of percussive elements during time-scaling.  The six best methods are separated by $<0.102$ OMOS, with a maximum difference of 0.263 for all methods. A two-sample t-test analysis ($\alpha=0.05$) of all OMOS shows the null hypothesis of equal means to be rejected in almost all cases when the absolute difference of mean OMOS is greater than 0.098.  Figure~\ref{fig:Masked_P_GRU} shows p-values for the t-tests that were unable to reject equal means.

The OMOQDE took approximately 15 hours to evaluate the 6000 files of the evaluation set, while the proposed networks took approximately 15 minutes using the same hardware.  The majority of this  improvement is due to the removal of time-frequency spreading when calculating PEAQ features. The elimination of alignment between reference and test signals is also beneficial as it removes an additional temporal manipulation before feature calculation.  While OMOQDE is a more accurate estimate of time-scaling quality, the proposed OMOQSE measures give a fast relative quality assessment, and provide a platform for future SE objective measures.

\section{Availability}
\label{sec:6}
The proposed CNN and BGRU-FT tools are available from github.com/zygurt/TSM.  This includes python scripts for feature generation, the proposed methods implemented in PyTorch and evaluation methods.  A bash script is also included to simplify use.

\section{Future Research}
\label{sec:7}
This study shows promise in non-invasive evaluation of the quality of TSM methods.  However, improvements can be made through input feature selection and exploring the use of phase derivatives or instantaneous frequency.  Generalization to unseen TSM methods and sound sources is also an area for future research.  More research needs to be conducted regarding duration invariant transformations that limit the networks ability to learn simple relationships such as the the duration of musical events within the signal to SMOS.  Additional attention could also be given to network architectures, such as Transformer networks \citep{Vaswani_2017_Attention}.  Pre-training using a large task related dataset could also be explored.

\section{Conclusion}
\label{sec:8}
Two single-ended objective measures for time-scaled audio are proposed with performance matching that of simple OMOQDE measures with reduced processing time.  CNN and BGRU-FT network architectures generate data-driven features from [MFCCs;$\textrm{D}$] inputs, which were are fed to an FCNN.  The networks are trained to the SMOS of the TSMDB.  The CNN achieves an $\overline{\bm{\mathcal{L}}}$ of 0.608 and  $\overline{\bm{\rho}}$ of 0.771, while BGRU-FT achieves an $\overline{\bm{\mathcal{L}}}$ of 0.576 and $\overline{\bm{\rho}}$ of 0.794.  The proposed measures are used to evaluate TSM methods, with estimates consistent with relative quality found in subjective testing.  Future work includes exploration of alternative features and other network architectures.

\bibliography{Manuscript.bib}

\end{document}

%% file: Table1.tex
\begin{table*}[ht]
\caption{Test Loss ($\bm{\mathcal{L}_{te}}$) and PCC ($\bm{\rho_{te}}$), mean RSME loss ($\bm{\overline{\mathcal{L}}}$) and range ($\bm{\Delta\mathcal{L}}$), mean PCC ($\bm{\overline{\rho}}$) and range ($\bm{\Delta\rho}$), median overall distance ($\bm{\widetilde{\mathcal{D}}}$) and minimum overall distance ($\bm{\text{min}(\mathcal{D})}$).  Best single-ended results in bold.}
\small
\centering
\begin{ruledtabular}
\begin{tabular}{ccccccccccccc}
\textbf{Ended} & \textbf{Network} & \textbf{Features} & \textbf{Hidden} & $\bm{\mathcal{L}_{te}}$ & $\bm{\rho_{te}}$ & $\bm{\overline{\mathcal{L}}}$ & $\bm{\Delta\mathcal{L}}$ & $\bm{\overline{\rho}}$ & $\bm{\Delta\rho}$ & $\bm{\widetilde{\mathcal{D}}}$ & $\bm{\text{min}(\mathcal{D})}$ \\
\hline
SE & BLSTM-FF & $|X|^2$ & 512 & 1.123 & 0.244 & 1.009 & 0.194 & 0.163 & 0.220 & 1.364 & 1.344 \\
\hline
SE & LLSTM-FF & $|X|^2$ & 512 & 1.064 & 0.262 & 0.989 & \textbf{0.135} & 0.171 & 0.251 & 1.350 & 1.322 \\
\hline
SE & CNN & [$|X|$;$\angle X$] & - & 0.942 & 0.484 & 0.850 & 0.188 & 0.523 & 0.099 & 1.039 & 0.998 \\
\hline
SE & CNN & $|X|$ & - & 0.944 & 0.553 & 0.745 & 0.339 & 0.674 & 0.205 & 0.991 & 0.904 \\
\hline
SE & LSTM-FF & [MFCCs;$\textrm{D}$] & 256 & 0.854 & 0.581 & 0.663 & 0.295 & 0.720 & 0.221 & 0.898 & 0.809 \\
\hline
SE & BLSTM-FF & [MFCCs;$\textrm{D}$] & 256 & 0.849 & 0.581 & 0.670 & 0.282 & 0.711 & 0.214 & 0.887 & 0.811 \\
\hline
SE & GRU-FT & [MFCCs;$\textrm{D}$] & 64 & 0.820 & 0.649 & 0.699 & 0.188 & 0.701 & \textbf{0.097} & 0.855 & 0.789 \\
\hline
SE & BGRU-FT & [MFCCs;$\textrm{D}$] & 64 & 0.778 & 0.675 & 0.611 & 0.287 & 0.770 & 0.179 & 0.809 & 0.735 \\
\hline
SE & CNN & [MFCCs;$\textrm{D}$] & - & 0.801 & 0.637 & 0.608 & 0.301 & 0.771 & 0.206 & 0.808 & 0.745 \\
\hline
SE & BGRU-FF & [MFCCs;$\textrm{D}$] & 256 & 0.784 & 0.667 & 0.622 & 0.248 & 0.762 & 0.153 & 0.789 & 0.727 \\
\hline
DE & FCNN & PEAQB & 3 & 0.691 & 0.749 & 0.668 & 0.054 & 0.719 & 0.075 & 0.756 & 0.731 \\
\hline
SE & BGRU-FT & [MFCCs;$\textrm{D}$] & 256 & \textbf{0.762} & \textbf{0.682} & \textbf{0.576} & 0.307 & \textbf{0.794} & 0.192 & \textbf{0.738} & \textbf{0.711} \\
\hline
DE & FCNN & PEAQB & 128 & 0.704 & 0.742 & 0.650 & 0.091 & 0.748 & 0.009 & 0.738 & 0.703 \\
\hline
DE & FCNN & To Test Incl Ref & 128 & 0.550 & 0.852 & 0.490 & 0.101 & 0.864 & 0.030 & 0.600 & 0.519 \\
\end{tabular}
\end{ruledtabular}
\label{tab:RMSE_PCC}
\end{table*}

%% file: Table2.tex
\begin{table*}
\caption{Mean OMOS for each class of file and overall result for the proposed CNN OMOQ. Means calculated for $\beta\neq 0.2257$ and $\beta\neq 1$.}
\small
\centering
\begin{ruledtabular}
\begin{tabular}{cccccccccccccccc}
 & \textbf{ES} & \textbf{FES} & \textbf{PV} & \textbf{WS} & \textbf{NMF} & \textbf{FPV} & \textbf{SPL} & \textbf{DIPL} & $\overline{\textbf{SPL}}$ & $\overline{\textbf{IPL}}$ & \textbf{IPL} & \textbf{HP} & $\overline{\textbf{uTVS}}$ & \textbf{uTVS} & \textbf{EL} \\
\hline
Music & 2.291 & 2.424 & 3.318 & 3.045 & 3.053 & 3.290 & 3.259 & 3.299 & 3.327 & 3.335 & 3.294 & 3.325 & 3.460 & \textbf{3.469} & 3.445 \\
\hline
Solo & 3.248 & 3.209 & 3.202 & 3.416 & 3.370 & 3.396 & 3.297 & 3.343 & 3.375 & 3.441 & 3.372 & \textbf{3.553} & 3.424 & 3.435 & 3.419 \\
\hline
Voice & 2.966 & 2.938 & 2.793 & 3.020 & 3.082 & 2.886 & 3.064 & 2.982 & 2.948 & 2.879 & 3.053 & 2.925 & 2.988 & 2.999 & \textbf{3.104} \\
\hline
Overall & 2.781 & 2.814 & 3.126 & 3.149 & 3.157 & 3.200 & 3.212 & 3.217 & 3.227 & 3.230 & 3.245 & 3.274 & 3.307 & 3.318 & \textbf{3.335} \\
\end{tabular}
\end{ruledtabular}
\label{tab:CNN_Eval}
\end{table*}

%% file: Table3.tex
\begin{table*}
\caption{Mean OMOS for each class of file and overall result for the proposed BGRU-FT network. Means calculated for $\beta\neq 0.2257$ and $\beta\neq 1$.}
\small
\centering
\begin{ruledtabular}
\begin{tabular}{cccccccccccccccc}
 & \textbf{ES} & \textbf{FES} & \textbf{PV} & \textbf{SPL} & \textbf{IPL} & $\overline{\textbf{SPL}}$ & $\overline{\textbf{IPL}}$ & \textbf{WS} & \textbf{HP} & \textbf{DIPL} & \textbf{NMF} & $\overline{\textbf{uTVS}}$ & \textbf{EL} & \textbf{uTVS} & \textbf{FPV} \\
\hline
Music & 2.378 & 2.511 & 2.723 & 2.711 & 2.764 & 2.733 & 2.764 & 2.741 & 2.699 & 2.787 & 2.720 & 2.890 & 2.899 & 2.901 & \textbf{3.016} \\
\hline
Solo & 2.925 & 2.947 & 2.891 & 2.917 & 2.917 & 2.943 & 2.976 & 2.978 & \textbf{3.035} & 2.978 & 2.988 & 2.983 & 2.988 & 3.005 & 3.011 \\
\hline
Voice & 2.503 & 2.468 & 2.323 & 2.350 & 2.329 & 2.345 & 2.334 & 2.472 & 2.492 & 2.443 & \textbf{2.591} & 2.526 & 2.544 & 2.532 & 2.444 \\
\hline
Overall & 2.580 & 2.629 & 2.653 & 2.664 & 2.680 & 2.680 & 2.699 & 2.732 & 2.738 & 2.741 & 2.762 & 2.809 & 2.819 & 2.822 & \textbf{2.843} \\
\end{tabular}
\end{ruledtabular}
\label{tab:BGRU_Eval}
\end{table*}

%% file: Manuscript.bbl
\begin{thebibliography}{32}
\def\enquote#1{``#1,''}
\def\plainquote#1{``#1''}
\expandafter\ifx\csname natexlab\endcsname\relax\def\natexlab#1{#1}\fi
\providecommand{\dourl}[1]{\href{http://#1}{\nolinkurl{#1}}}
\providecommand{\bibinfo}[2]{#2}
\providecommand{\noopsort}[1]{}
\providecommand{\switchargs}[2]{#2#1}
  \def\eatspace #1{#1}

\bibitem[{Beerends \emph{et~al.}(2013)Beerends, Schmidmer, Berger, Obermann,
  Ullmann, Pomy, and Keyhl}]{Beerends_2013_POLQA}
\bibinfo{author}{Beerends, J.~G.}, \bibinfo{author}{Schmidmer, C.},
  \bibinfo{author}{Berger, J.}, \bibinfo{author}{Obermann, M.},
  \bibinfo{author}{Ullmann, R.}, \bibinfo{author}{Pomy, J.},  and
  \bibinfo{author}{Keyhl, M.} (\textbf{\bibinfo{year}{2013}}).
  \enquote{\bibinfo{title}{Perceptual objective listening quality assessment
  (polqa), the third generation itu-t standard for end-to-end speech quality
  measurement part i—temporal alignment}} \bibinfo{journal}{Journal of the
  Audio Engineering Society} \textbf{61}(6), \bibinfo{pages}{366--384}.

\bibitem[{Chinen \emph{et~al.}(2020)Chinen, Lim, Skoglund, Gureev, O'Gorman,
  and Hines}]{Chinen_2020}
\bibinfo{author}{Chinen, M.}, \bibinfo{author}{Lim, F.~S.},
  \bibinfo{author}{Skoglund, J.}, \bibinfo{author}{Gureev, N.},
  \bibinfo{author}{O'Gorman, F.},  and \bibinfo{author}{Hines, A.}
  (\textbf{\bibinfo{year}{2020}}). \enquote{\bibinfo{title}{Visqol v3: An open
  source production ready objective speech and audio metric}}
  \bibinfo{journal}{arXiv preprint arXiv:2004.09584} .

\bibitem[{Cho \emph{et~al.}(2014)Cho, Van~Merri{\"e}nboer, Gulcehre, Bahdanau,
  Bougares, Schwenk, and Bengio}]{Cho_2014_GRU}
\bibinfo{author}{Cho, K.}, \bibinfo{author}{Van~Merri{\"e}nboer, B.},
  \bibinfo{author}{Gulcehre, C.}, \bibinfo{author}{Bahdanau, D.},
  \bibinfo{author}{Bougares, F.}, \bibinfo{author}{Schwenk, H.},  and
  \bibinfo{author}{Bengio, Y.} (\textbf{\bibinfo{year}{2014}}).
  \enquote{\bibinfo{title}{Learning phrase representations using rnn
  encoder-decoder for statistical machine translation}} \bibinfo{journal}{arXiv
  preprint arXiv:1406.1078} .

\bibitem[{Damsk{\"a}gg and V{\"a}lim{\"a}ki(2017)}]{Damskagg_2017}
\bibinfo{author}{Damsk{\"a}gg, E.},  and \bibinfo{author}{V{\"a}lim{\"a}ki, V.}
  (\textbf{\bibinfo{year}{2017}}). \enquote{\bibinfo{title}{Audio time
  stretching using fuzzy classification of spectral bins}}
  \bibinfo{journal}{Applied Sciences} \textbf{7}(12), \bibinfo{pages}{1293}.

\bibitem[{Davis and Mermelstein(1980)}]{Davis_1980_MFCC}
\bibinfo{author}{Davis, S.},  and \bibinfo{author}{Mermelstein, P.}
  (\textbf{\bibinfo{year}{1980}}). \enquote{\bibinfo{title}{Comparison of
  parametric representations for monosyllabic word recognition in continuously
  spoken sentences}} \bibinfo{journal}{IEEE transactions on acoustics, speech,
  and signal processing} \textbf{28}(4), \bibinfo{pages}{357--366}.

\bibitem[{Driedger \emph{et~al.}(2014)Driedger, Muller, and
  Ewert}]{Driedger_Muller_Ewert_2014}
\bibinfo{author}{Driedger, J.}, \bibinfo{author}{Muller, M.},  and
  \bibinfo{author}{Ewert, S.} (\textbf{\bibinfo{year}{2014}}).
  \enquote{\bibinfo{title}{Improving time-scale modification of music signals
  using harmonic-percussive separation}} \bibinfo{journal}{IEEE Signal
  Processing Letters} \textbf{21}(1), \bibinfo{pages}{105--109}.

\bibitem[{Falk and Chan(2006)}]{Falk_2006}
\bibinfo{author}{Falk, T.~H.},  and \bibinfo{author}{Chan, W.-Y.}
  (\textbf{\bibinfo{year}{2006}}). \enquote{\bibinfo{title}{Single-ended speech
  quality measurement using machine learning methods}} \bibinfo{journal}{IEEE
  Transactions on Audio, Speech, and Language Processing} \textbf{14}(6),
  \bibinfo{pages}{1935--1947}.

\bibitem[{Fierro and V{\"a}lim{\"a}ki(2020)}]{Fierro_2020}
\bibinfo{author}{Fierro, L.},  and \bibinfo{author}{V{\"a}lim{\"a}ki, V.}
  (\textbf{\bibinfo{year}{2020}}). \enquote{\bibinfo{title}{Towards objective
  evaluation of audio time-scale modification methods}} in
  \emph{\bibinfo{booktitle}{Proceedings of the 17th Sound and Music Computing
  Conference}}, pp. \bibinfo{pages}{457--462}.

\bibitem[{Gamper \emph{et~al.}(2019)Gamper, Reddy, Cutler, Tashev, and
  Gehrke}]{Gamper_2019}
\bibinfo{author}{Gamper, H.}, \bibinfo{author}{Reddy, C.~K.},
  \bibinfo{author}{Cutler, R.}, \bibinfo{author}{Tashev, I.~J.},  and
  \bibinfo{author}{Gehrke, J.} (\textbf{\bibinfo{year}{2019}}).
  \enquote{\bibinfo{title}{Intrusive and non-intrusive perceptual speech
  quality assessment using a convolutional neural network}} in
  \emph{\bibinfo{booktitle}{2019 IEEE Workshop on Applications of Signal
  Processing to Audio and Acoustics (WASPAA)}}, \bibinfo{organization}{IEEE},
  pp. \bibinfo{pages}{85--89}.

\bibitem[{Hochreiter and Schmidhuber(1997)}]{Hochreiter_1997_LSTM}
\bibinfo{author}{Hochreiter, S.},  and \bibinfo{author}{Schmidhuber, J.}
  (\textbf{\bibinfo{year}{1997}}). \enquote{\bibinfo{title}{Long short-term
  memory}} \bibinfo{journal}{Neural computation} \textbf{9}(8),
  \bibinfo{pages}{1735--1780}.

\bibitem[{Huber and Kollmeier(2006)}]{Huber_2006}
\bibinfo{author}{Huber, R.},  and \bibinfo{author}{Kollmeier, B.}
  (\textbf{\bibinfo{year}{2006}}). \enquote{\bibinfo{title}{Pemo-q—a new
  method for objective audio quality assessment using a model of auditory
  perception}} \bibinfo{journal}{IEEE Transactions on audio, speech, and
  language processing} \textbf{14}(6), \bibinfo{pages}{1902--1911}.

\bibitem[{ITU-T(2001)}]{ITU_BS1387_PEAQ}
\bibinfo{author}{ITU-T} (\textbf{\bibinfo{year}{2001}}).
  \enquote{\bibinfo{title}{Itu-r bs. 1387-1: Method for objective measurements
  of perceived audio quality}} \bibinfo{type}{Technical Report}.

\bibitem[{Karrer \emph{et~al.}(2006)Karrer, Lee, and
  Borchers}]{Karrer_Lee_Borchers_2006}
\bibinfo{author}{Karrer, T.}, \bibinfo{author}{Lee, E.},  and
  \bibinfo{author}{Borchers, J.} (\textbf{\bibinfo{year}{2006}}).
  \enquote{\bibinfo{title}{{PhaVoRIT}: A phase vocoder for real-time
  interactive time-stretching}} \bibinfo{type}{Technical Report}.

\bibitem[{Kim(2005)}]{Kim_2005}
\bibinfo{author}{Kim, D.-S.} (\textbf{\bibinfo{year}{2005}}).
  \enquote{\bibinfo{title}{Anique: An auditory model for single-ended speech
  quality estimation}} \bibinfo{journal}{IEEE Transactions on Speech and Audio
  Processing} \textbf{13}(5), \bibinfo{pages}{821--831}.

\bibitem[{Laroche and Dolson(1999)}]{Laroche_Dolson_1999_IPL}
\bibinfo{author}{Laroche, J.},  and \bibinfo{author}{Dolson, M.}
  (\textbf{\bibinfo{year}{1999}}). \enquote{\bibinfo{title}{Improved phase
  vocoder time-scale modification of audio}} \bibinfo{journal}{IEEE
  Transactions on Speech and Audio Processing} \textbf{7}(3),
  \bibinfo{pages}{323--332}.

\bibitem[{LeCun \emph{et~al.}(2015)LeCun, Bengio, and Hinton}]{Lecun_2015}
\bibinfo{author}{LeCun, Y.}, \bibinfo{author}{Bengio, Y.},  and
  \bibinfo{author}{Hinton, G.} (\textbf{\bibinfo{year}{2015}}).
  \enquote{\bibinfo{title}{Deep learning}} \bibinfo{journal}{nature}
  \textbf{521}(7553), \bibinfo{pages}{436--444}.

\bibitem[{Loshchilov and Hutter(2017)}]{AdamW_2017}
\bibinfo{author}{Loshchilov, I.},  and \bibinfo{author}{Hutter, F.}
  (\textbf{\bibinfo{year}{2017}}). \enquote{\bibinfo{title}{Decoupled weight
  decay regularization}} \bibinfo{journal}{arXiv preprint arXiv:1711.05101} .

\bibitem[{Malfait \emph{et~al.}(2006)Malfait, Berger, and
  Kastner}]{Malfait_2006}
\bibinfo{author}{Malfait, L.}, \bibinfo{author}{Berger, J.},  and
  \bibinfo{author}{Kastner, M.} (\textbf{\bibinfo{year}{2006}}).
  \enquote{\bibinfo{title}{P. 563—the itu-t standard for single-ended speech
  quality assessment}} \bibinfo{journal}{IEEE Transactions on Audio, Speech,
  and Language Processing} \textbf{14}(6), \bibinfo{pages}{1924--1934}.

\bibitem[{Nicolson \emph{et~al.}(2018)Nicolson, Hanson, Lyons, and
  Paliwal}]{Nicolson_2018}
\bibinfo{author}{Nicolson, A.}, \bibinfo{author}{Hanson, J.},
  \bibinfo{author}{Lyons, J.},  and \bibinfo{author}{Paliwal, K.}
  (\textbf{\bibinfo{year}{2018}}). \enquote{\bibinfo{title}{Spectral subband
  centroids for robust speaker identification using marginalization-based
  missing feature theory}} \bibinfo{journal}{International Journal of Signal
  Processing Systems} \textbf{6}(1), \bibinfo{pages}{12--16},
  \dodoi{10.18178/ijsps.6.1.12-16}.

\bibitem[{Portnoff(1976)}]{Portnoff_1976}
\bibinfo{author}{Portnoff, M.} (\textbf{\bibinfo{year}{1976}}).
  \enquote{\bibinfo{title}{Implementation of the digital phase vocoder using
  the fast {F}ourier transform}} \bibinfo{journal}{IEEE Transactions on
  Acoustics, Speech, And Signal Processing} \textbf{24}(3),
  \bibinfo{pages}{243--248}.

\bibitem[{Rix \emph{et~al.}(2001)Rix, Beerends, Hollier, and
  Hekstra}]{Rix_2001_PESQ}
\bibinfo{author}{Rix, A.~W.}, \bibinfo{author}{Beerends, J.~G.},
  \bibinfo{author}{Hollier, M.~P.},  and \bibinfo{author}{Hekstra, A.~P.}
  (\textbf{\bibinfo{year}{2001}}). \enquote{\bibinfo{title}{Perceptual
  evaluation of speech quality (pesq)-a new method for speech quality
  assessment of telephone networks and codecs}} in
  \emph{\bibinfo{booktitle}{2001 IEEE International Conference on Acoustics,
  Speech, and Signal Processing. Proceedings (Cat. No. 01CH37221)}},
  \bibinfo{organization}{IEEE}, Vol. 2, pp. \bibinfo{pages}{749--752}.

\bibitem[{Roberts and Paliwal(2019)}]{Roberts_2019_FESOLA}
\bibinfo{author}{Roberts, T.},  and \bibinfo{author}{Paliwal, K.~K.}
  (\textbf{\bibinfo{year}{2019}}). \enquote{\bibinfo{title}{Time-scale
  modification using fuzzy epoch-synchronous overlap-add ({FESOLA})}} in
  \emph{\bibinfo{booktitle}{2019 IEEE Workshop on Applications of Signal
  Processing to Audio and Acoustics}}, \bibinfo{organization}{IEEE}, pp.
  \bibinfo{pages}{31--34}.

\bibitem[{Roberts and Paliwal(2020{\natexlab{a}})}]{Roberts_2020_OMOQ}
\bibinfo{author}{Roberts, T.},  and \bibinfo{author}{Paliwal, K.~K.}
  (\textbf{\bibinfo{year}{2020}}{\natexlab{a}}). \enquote{\bibinfo{title}{An
  objective measure of quality for time-scale modification of audio}}
  \bibinfo{journal}{arXiv preprint arXiv:2006.06153} \bibinfo{note}{{Under
  Review with JASA}}.

\bibitem[{Roberts and Paliwal(2020{\natexlab{b}})}]{Roberts_2020_SMOS}
\bibinfo{author}{Roberts, T.},  and \bibinfo{author}{Paliwal, K.~K.}
  (\textbf{\bibinfo{year}{2020}}{\natexlab{b}}). \enquote{\bibinfo{title}{A
  time-scale modification dataset with subjective quality labels}}
  \bibinfo{journal}{The Journal of the Acoustical Society of America}
  \textbf{145}(5), \bibinfo{pages}{3095--3103}.

\bibitem[{Roma \emph{et~al.}(2019)Roma, Green, and
  Tremblay}]{Roma_Green_Tremblay_2019}
\bibinfo{author}{Roma, G.}, \bibinfo{author}{Green, O.},  and
  \bibinfo{author}{Tremblay, P.} (\textbf{\bibinfo{year}{2019}}).
  \enquote{\bibinfo{title}{Time scale modification of audio using non-negative
  matrix factorization}} in \emph{\bibinfo{booktitle}{Proc. of the 22nd Int.
  Conference on Digital Audio Effects (DAFx-19)}},
  \bibinfo{address}{Birmingham, UK}, pp. \bibinfo{pages}{1--6}.

\bibitem[{Rudresh \emph{et~al.}(2018)Rudresh, Vasisht, Vijayan, and
  Seelamantula}]{Rudresh_2018}
\bibinfo{author}{Rudresh, S.}, \bibinfo{author}{Vasisht, A.},
  \bibinfo{author}{Vijayan, K.},  and \bibinfo{author}{Seelamantula, C.~S.}
  (\textbf{\bibinfo{year}{2018}}). \enquote{\bibinfo{title}{Epoch-synchronous
  overlap-add (esola) for time-and pitch-scale modification of speech signals}}
  \bibinfo{journal}{arXiv preprint arXiv:1801.06492}
  \bibinfo{note}{Unpublished}.

\bibitem[{Schuster and Paliwal(1997)}]{Schuster_1997}
\bibinfo{author}{Schuster, M.},  and \bibinfo{author}{Paliwal, K.~K.}
  (\textbf{\bibinfo{year}{1997}}). \enquote{\bibinfo{title}{Bidirectional
  recurrent neural networks}} \bibinfo{journal}{IEEE transactions on Signal
  Processing} \textbf{45}(11), \bibinfo{pages}{2673--2681}.

\bibitem[{Sharma \emph{et~al.}(2017)Sharma, Potadar, Chetupalli, and
  Sreenivas}]{Sharma_2017}
\bibinfo{author}{Sharma, N.}, \bibinfo{author}{Potadar, S.},
  \bibinfo{author}{Chetupalli, S.~R.},  and \bibinfo{author}{Sreenivas, T.}
  (\textbf{\bibinfo{year}{2017}}). \enquote{\bibinfo{title}{Mel-scale sub-band
  modelling for perceptually improved time-scale modification of speech and
  audio signals}} in \emph{\bibinfo{booktitle}{2017 Twenty-third National
  Conference on Communications (NCC)}}, \bibinfo{organization}{IEEE}, pp.
  \bibinfo{pages}{1--5}.

\bibitem[{Thiede \emph{et~al.}(2000)Thiede, Treurniet, Bitto, Schmidmer,
  Sporer, Beerends, and Colomes}]{Thiede_2000}
\bibinfo{author}{Thiede, T.}, \bibinfo{author}{Treurniet, W.~C.},
  \bibinfo{author}{Bitto, R.}, \bibinfo{author}{Schmidmer, C.},
  \bibinfo{author}{Sporer, T.}, \bibinfo{author}{Beerends, J.~G.},  and
  \bibinfo{author}{Colomes, C.} (\textbf{\bibinfo{year}{2000}}).
  \enquote{\bibinfo{title}{Peaq-the itu standard for objective measurement of
  perceived audio quality}} \bibinfo{journal}{Journal of the Audio Engineering
  Society} \textbf{48}(1/2), \bibinfo{pages}{3--29}.

\bibitem[{Vaswani \emph{et~al.}(2017)Vaswani, Shazeer, Parmar, Uszkoreit,
  Jones, Gomez, Kaiser, and Polosukhin}]{Vaswani_2017_Attention}
\bibinfo{author}{Vaswani, A.}, \bibinfo{author}{Shazeer, N.},
  \bibinfo{author}{Parmar, N.}, \bibinfo{author}{Uszkoreit, J.},
  \bibinfo{author}{Jones, L.}, \bibinfo{author}{Gomez, A.~N.},
  \bibinfo{author}{Kaiser, {\L}.},  and \bibinfo{author}{Polosukhin, I.}
  (\textbf{\bibinfo{year}{2017}}). \enquote{\bibinfo{title}{Attention is all
  you need}} in \emph{\bibinfo{booktitle}{Advances in neural information
  processing systems}}, pp. \bibinfo{pages}{5998--6008}.

\bibitem[{Verhelst and Roelands(1993)}]{Verhelst_Roelands_1993}
\bibinfo{author}{Verhelst, W.},  and \bibinfo{author}{Roelands, M.}
  (\textbf{\bibinfo{year}{1993}}). \enquote{\bibinfo{title}{An overlap-add
  technique based on waveform similarity ({WSOLA}) for high quality time-scale
  modification of speech}} in \emph{\bibinfo{booktitle}{Proceedings of ICASSP
  '93}}, \bibinfo{organization}{IEEE}, Vol. 2, pp. \bibinfo{pages}{554--557}.

\bibitem[{{Zplane Development}()}]{elastique}
\bibinfo{author}{{Zplane Development}}.
  \plainquote{\bibinfo{title}{\`{E}lastique time stretching \& pitch shifting
  sdks (version 3.2.5) [computer program]}}
  \dourl{http://licensing.zplane.de/technology\#elastique},
  \bibinfo{note}{(Last viewed October 31, 2019)}.

\end{thebibliography}
